\documentstyle[11pt]{article}

\def\thefootnote{\fnsymbol{footnote}}
\catcode`\@=11
\def\beq{\begin{equation}}
\def\eeq{\end{equation}}
\def\barr{\begin{eqnarray}}
\def\beqa{\begin{eqnarray}}
\def\earr{\end{eqnarray}}
\def\eeqa{\end{eqnarray}}
\def\winf{W_{1+\infty}\ }
\def\u1{\widehat{U(1)}}
\def\v{V\,}
\def\w{W\,}
\def\vb{{\overline V}\,}
\def\wb{{\overline W}\,}

\newcommand{\eq}[1]{Eq.~(\ref{#1})}

\newcommand{\nl}{\nonumber \\}

\newcommand{\EQ}{\begin{equation}} \newcommand{\EN}{\end{equation}}
\newcommand{\bea}{\begin{eqnarray}}
\newcommand{\ena}{\end{eqnarray}}

\newcommand{\NP}[1]{{\it Nucl.\ Phys.\ }{\bf #1}}
\newcommand{\PL}[1]{{\it Phys.\ Lett.\ }{\bf #1}}

\newcommand{\CMP}[1]{{\it Comm.\ Math.\ Phys.\ }{\bf #1}}
\newcommand{\PR}[1]{{\it Phys.\ Rev.\ }{\bf #1}}
\newcommand{\PRL}[1]{{\it Phys.\ Rev.\ Lett.\ }{\bf #1}}

\newcommand{\IJMP}[1]{{\it Int.\ Jour.\ Mod.\ Phys.\ }{\bf #1}}
\renewcommand{\thefootnote}{\fnsymbol{footnote}}
\begin{document}
\begin{titlepage}
\begin{center}
\hfill DFTT 21/96 \\
\hfill June 1996
\vskip 0.8cm
{\Large \bf The extended conformal theory of
Luttinger systems \footnote{Talk given by S. Sciuto at the
{\it II} International Sakharov Conference on
Physics'', Moscow, Russia,
{}~~20--24 May 96.}}
\vskip 0.8cm
Marialuisa~FRAU,~~Stefano~SCIUTO \\
\vskip 0.3cm
{\em      Dipartimento di Fisica Teorica,
          Universit\`a di Torino,\\
             and I.N.F.N. Sezione di Torino \\
          Via P. Giuria 1, I-10125 Torino, Italy}
\vskip 0.6cm
Alberto LERDA \\

\vskip 0.3cm
{\em       Dipartimento di Scienze e Tecnologie Avanzate \footnote{
           II Facolt\'a di Scienze M.F.N., Universit\`a di Torino
           (sede di Alessandria), Italy.} and\\
           Dipartimento di Fisica Teorica, Universit\`a di Torino,\\
           and I.N.F.N., Sezione di Torino \\
           Via P.Giuria 1, I-10125 Torino, Italy}
\vskip 0.6cm
Guillermo~R.~ZEMBA \\
\vskip 0.3cm
{\em       C.N.E.A. , Centro At\'omico Bariloche and \\
		   Instituto Balseiro, Universidad Nacional de Cuyo\\
           8400 - San Carlos de Bariloche ~(R\'{\i}o Negro),
           Argentina}
\end{center}
\vskip 1cm
\begin{abstract}
\noindent
We describe the recently introduced method of algebraic
bosonization of the $(1+1)$-dimensional Luttinger systems by
discussing in detail the specific case of the Calogero-Sutherland
model,
and mentioning the hard-core Bose gas.
We also compare our findings with the exact Bethe Ansatz results.
\end{abstract}
\vfill \end{titlepage}

\pagenumbering{arabic}
\renewcommand{\thefootnote}{\arabic{footnote}}
\setcounter{footnote}{0}
\setcounter{page}{1}

It has been known for some time \cite {hald1,hald} that the
one-dimensional gapless fermionic systems solvable by
Bethe Ansatz, have leading finite
size corrections that are described by the Luttinger model
\cite{lutt}.
For this reason, such systems have been called Luttinger
systems \cite{hald}. More recently \cite{fermi}, it has become
clear that the Luttinger model is simply a $c=1$
conformal field theory.
In this talk we will show that also the higher order
corrections to the thermodinamic limit of the Luttinger
systems have a simple algebraic interpretation \cite{flsz}, namely
they
are described by an extended conformal field theory based on
the $\winf \times {\overline \winf}$ algebra \cite{shen,kac1},
which is an infinite extension of the conformal Virasoro algebra.
We illustrate this fact on some specific examples, starting
from the Calogero-Sutherland model.

The Calogero-Sutherland \cite{cal}
model describes $N$ non-relativistic
spinless fermions moving on a circle of length $L$ with
hamiltonian
$$
h=-\sum_{j=1}^N \frac{\partial^2}{\partial x_j^2}
+g\,\frac{\pi^2}{L^2}\sum_{j<k}
\frac{1}{\sin^2 (\pi(x_j-x_k)/L) }~~~,
$$
The second quantized operator corresponding to $h$ is
the sum of the kinetic term
\beq
H_0 = \left(\frac{2\pi}{L}\right)^2\sum_{n=-\infty}^\infty
n^2\,\psi_n^{\dagger}\,\psi_n~~~,
\label{h0cs}
\eeq
and the interaction term \cite{clz}
\beq
H_I =- g \,\frac{\pi^2}{L^2} \sum_{l,n,m=-\infty}^\infty |l|
{}~\psi_{m+l}^{\dagger}\,\psi_{n-l}^{\dagger}\,
\psi_n \,\psi_m ~~~,
\label{hics}
\eeq
where $\psi_n$ are fermionic oscillators of momentum
$k_n=\left(2\pi/L\right)n$ ($n \in {\bf Z}$ if $N$
is odd and $n \in {\bf Z}' \equiv {\bf Z}+1/2$ if $N$ is even),
satisfying standard anticommutation relations.

The hamiltonian $H_0$ describes $N$ free fermions
whose ground state is
$$
|\Omega\rangle = \psi^\dagger_{-n_F} \dots \psi^\dagger_{n_F}
|0\rangle~~~~,
$$
with $n_F = (N-1)/2$ .
To describe the long-distance properties
of the system, it is enough to consider
only those oscillators near the Fermi points $\pm n_F$.
Thus, the effective theory can be conveniently written in terms
of the shifted operators
\beq
a_r\equiv\psi_{n_F+r+\frac{1}{2}}~~~,~~~
b_r\equiv\psi_{-n_F-r-\frac{1}{2}}~~~~.
\label{arbr}
\eeq
The oscillators $a_r$ ($b_r$) describe
small fluctuations of momentum
$2\pi r/L$ ($-2\pi r/L$)
relative to the right (left) Fermi point.
The half-integer index $r$ is allowed to vary only in a
finite range between $-\Lambda$ and $+\Lambda$,
where the bandwidth cut-off $\Lambda$ is such that
$\Lambda \ll n_F$. Furthermore,
\bea
a_r|\Omega\rangle = b_r|\Omega\rangle =
a_{-r}^\dagger|\Omega\rangle = b_{-r}^\dagger|\Omega\rangle = 0
{}~~~~{\rm for}~~~r > 0~~~.
\label{abvac1}
\ena

Using these shifted oscillators, the effective hamiltonian
corresponding to
the kinetic part $H_0$ reads
\bea
{\cal H}_0 &=&
\left(\frac{2\pi}{L}\right)^2\
\sum_{r=-\Lambda}^{\Lambda}\
\left(n_F+r + \frac{1}{2}\right)^2 \,
\left(:a^\dagger_r\,a_r:+:b^\dagger_r\,b_r:\right)
\nl
&=& \left(2\pi\rho_0\right)^2 \sum_{r=-\Lambda}^\Lambda
\left( \frac{1}{4} + \frac{r}{N} +
\frac{r^2}{N^2} \right)
\left(:a^{\dagger}_r \,a_r:+:b^{\dagger}_r \,b_r:\right)\ ,
\label{kindue}
\ena
where $\rho_0 =N/L$ is the density, which is held fixed in
the thermodynamic limit.

If we keep only the leading $1/N$-term of ${\cal H}_0$,
{\it i.e.} if we linearize
the dispersion relation around the Fermi points, we can safely let
$\Lambda \to \infty$. Indeed, the spurious states introduced
by removing the band-width cut-off, have very high energy, and thus
can be neglected in the effective theory.
This procedure is the same that was used originally
to map the Tomonaga model into the Luttinger
model \cite{fermi}.
However, if we
want to keep also the higher order terms in the $1/N$-expansion,
we are allowed to extend the sum over $r$
in \eq{kindue} up to infinity, provided that at the same time we
keep $\Lambda \ll N$ and restrict the Hilbert space
to states with relative momentum bounded by $\Lambda$ \cite{flsz}.

The effective hamiltonian ${\cal H}_I$ associated to the
interaction term $H_I$ of \eq{hics}
can be treated in a similar way; the
detailed derivation of ${\cal H}_I$, which requires a suitable
regularization procedure and a careful dealing of normal ordering
effects in the backward scattering contribution, is presented in
Ref. \cite{flsz}. Here we simply write the complete
final result, namely
\beq
{\cal H} = {\cal H}_0 + {\cal H}_I = \left(2\pi \rho_0\right)^2
\sum_{k=0}^2 \frac{1}{N^k}\,{\cal H}_{(k)}
\label{seriesc}
\eeq
where
\beq
{\cal H}_{(0)} \ =\ \frac{1}{4}\,(1+g)\sum_{r \in {\bf Z}'}
\left( : a^{\dagger}_r \,a_r: +:b^{\dagger}_r\, b_r:\right)~~~,
\label{hc0}
\eeq
\bea
{\cal H}_{(1)} &=& \left(1+\frac{g}{2}\right)
\sum_{r \in {\bf Z}} r
\left( : a^{\dagger}_r \,a_r:+:b^{\dagger}_r\, b_r:\right)\nl
&& + \  \frac{g}{2}\sum_{\ell \in {\bf Z}}\ \sum_{r,s \in {\bf Z}'}
:a^\dagger_{r-\ell}\,a_r:\,:b^\dagger_{s-\ell}\,b_s:~~~,
\label{hc1}
\ena
and
\bea
{\cal H}_{(2)} &=& \sum_{r \in {\bf Z}'}
\left[r^2 + \frac{g}{4}\left(r^2-\frac{1}{4}\right)\right]
\left(:a^{\dagger}_r \,a_r:+:b^{\dagger}_r\,b_r:\right)
\nl
&&-\frac{g}{4} \sum_{\ell \in {\bf Z}}\ \sum_{r,s \in {\bf Z}'}
|\ell|\left(:a^\dagger_{r-\ell}\,a_r:\,:a^\dagger_{s+\ell}\,a_s:
+:b^\dagger_{r+\ell}\,b_r:\,:b^\dagger_{s-\ell}\,b_s: \right.\nl
&&~~~~~~~~~~~~~~~~
+\left.2\,:a^\dagger_{r-\ell}\,a_r:\,:b^\dagger_{s-\ell}\,b_s:
\right)
\label{hc2}\\
&&+\frac{g}{2} \sum_{\ell \in {\bf Z}}\sum_{r,s \in {\bf Z}'}
(r+s-\ell):a^\dagger_{r-\ell}\,a_r:\,:b^\dagger_{s-\ell}\,b_s: ~~~.
\nonumber
\ena
Notice that there are no contributions to ${\cal H}$ to order $1/N^3$
and
higher.

We now show that there is an elegant algebraic structure underlying
the
effective hamiltonian (\ref{seriesc}).
To see this, we first introduce the fermionic bilinear operators
\barr
\v^0_\ell &=& \sum_{r \in {\bf Z}'}: a^\dagger_{r-\ell}\, a_r :
{}~~~,\nl
\v^1_\ell &=& \sum_{r \in {\bf Z}'} \left(\,r-\frac{\ell}{2}\,
\right) : a^\dagger_{r-\ell}\, a_r :~~~,\label{fockw}\\
\v^2_\ell &=& \sum_{r \in {\bf Z}'} \left(\,r^2 -\ell r +
\frac{\ell^2}{6} + \frac{1}{12}\,
\right) :  a^\dagger_{r-\ell}\, a_r :~~~,
\nonumber
\earr
and $\vb_\ell^0$, $\vb_\ell^1$ and $\vb_\ell^2$ defined as above
with $: a^\dagger_{r-\ell}\, a_r :$ replaced by
$: b^\dagger_{r-\ell}\, b_r :$. Then, the operators
${\cal H}_{(k)}$ of Eqs. (\ref{hc0}--\ref{hc2}) become
\beq
{\cal H}_{(0)} =  \frac{1}{4} (1+g) \left(\v^0_0+
\vb^0_0\right)~~~,
\label{hcv0}
\eeq
\beq
{\cal H}_{(1)} = \left(1+\frac{g}{2}\right) \left(\v^1_0+
\vb^1_0\right)+\frac{g}{2} \sum_{\ell=-\infty}^{\infty}
\v^0_{\ell}\, \vb^0_{\ell}~~~,
\label{hcv1}
\eeq
\bea
{\cal H}_{(2)} &=&
\left(1+\frac{g}{4}\right) \left(\v^2_0+\vb^2_0\right)
-\frac{1}{12} (1+g) \left(\v^0_0+\vb^0_0\right) \nl
&& - \frac{g}{4}\sum_{\ell=-\infty}^{\infty} |\ell |
\left( \v^0_{\ell}\,\v^0_{-\ell}+
\vb^0_{-\ell} \,\vb^0_{\ell}+
2\,\v^0_{\ell} \,\vb^0_{\ell}\right)\nl
&& + \frac{g}{2}
\sum_{\ell=-\infty}^{\infty} \left( \v^1_{\ell}\,\vb^0_{\ell}+
\v^0_{\ell} \,\vb^1_{\ell} \right) ~~~.
\label{hcv2}
\ena
The operators introduced
in Eqs. (\ref{fockw}) are the lowest generators of the
infinite dimensional $\winf$algebra \cite{shen,kac1} whose general
form is
\beq
\left[\ V^i_\ell, V^j_m\ \right] = (j\ell-im)\, V^{i+j-1}_{\ell+m}
+q(i,j,\ell,m)\,V^{i+j-3}_{\ell+m}
+\cdots +c\ \delta^{ij}\delta_{\ell+m,0}\, d(i,\ell) ~~~.
\label{walg}
\eeq
Here $q(i,j,\ell,m)$ and $d(i,\ell)$
are polynomial structure constants,
$c$ is the central charge,
and the dots denote a finite number of terms involving the operators
$V^{i+j-2k}_{\ell+m}\ $.
In our case $c=1$, and the commutation relations relevant for our
purposes
are
\barr
\left[\ V^0_\ell,V^0_m\ \right] & = &  \ell\ \delta_{\ell+m,0} ~~~,
\label{walg0} \\
\left[\ V^1_\ell, V^0_m\ \right] & = & -m\ V^0_{\ell+m} ~~~,
\label{walg01}\\
\left[\ V^1_\ell, V^1_m\ \right] & = & (\ell-m)V^1_{\ell+m} +
\frac{1}{12}\ell(\ell^2-1) \delta_{\ell+m,0}~~~,
\label{walg1}\\
\left[\ V^2_\ell, V^0_m\ \right] &=& -2m\ V^1_{\ell+m}~~~,
\label{walg20}\\
\left[\ V^2_\ell, V^1_m\ \right] &=& (\ell-2m)\ V^2_{\ell+m} -
   \frac{1}{6}\left(m^3-m\right) V^0_{\ell+m}~~~.
\label{walg21}
\earr
{}From Eqs. (\ref{walg0}) and (\ref{walg1}) we see that
the generators $V^0_\ell$ satisfy an
Abelian Kac-Moody algebra, while the generators $V^1_\ell$
close a $c=1$ Virasoro algebra.
The operators $\vb^i_\ell$ satisfy the same algebra (\ref{walg})
and commute with the $\v^i_\ell$'s.

The $c=1$ $\winf$algebra can be also realized by {\it bosonic}
operators, through a generalized Sugawara construction \cite{kac1}.
In fact, if one introduces the right and left moving modes,
$\alpha_\ell$ and ${\overline \alpha}_\ell$, of a free compactified
boson (with the usual commutation relations and canonical normal
ordering),
one can check that the commutation relations (\ref{walg})
are satisfied by defining $\v^i_\ell$ (we only write the expressions
for $i=0,1,2$) as
\barr
\v^0_\ell &=& \alpha_\ell ~~~,
\label{mod0}\\
\v^1_\ell &=& {\frac{1}{2}} \sum_{n= -\infty}^{\infty}
:\, \alpha_{n}\,\alpha_{\ell-n}\,
:~~~,\label{mod1}\\
\v^2_\ell &=& {\frac{1}{3}} \sum_{n, m = -\infty}^{\infty}
:\, \alpha_{n}\,\alpha_m\, \alpha_{\ell-n-m}\,:~~~,
\label{mod2}
\earr
and analogously the $\vb^i_\ell$ in terms of
${\overline \alpha}_\ell$.

The major advantage for choosing the basis of the $\winf \times
{\overline \winf}$ operators is that, once the algebraic content
of the theory has been established in the free fermionic picture,
other bosonic realizations of the {\it same} algebra can be used,
and the free value of the compactification radius of the boson
can be chosen to diagonalize the total hamiltonian.
This is the reason for calling this procedure
{\it algebraic bosonization} \cite{flsz}.

In the fermionic description it is easy to see that the highest
weight states of the $\winf \times {\overline \winf}$ algebra
are obtained by adding $\Delta N$ particles to the
ground state $|\Omega \rangle$, and by moving $d$
particles from the left to the right Fermi point; they are
denoted by $|\Delta N,d\rangle_0$.
The descendant states,
$$
|\Delta N , d ; \{k_i\},\{{\overline k}_j\}
\rangle_0 \ =
\v^0_{-k_1} \dots \v^0_{-k_r} \vb^0_{-{\overline k}_1}
\dots \vb^0_{-{\overline k}_s}
|\Delta N , d \rangle_0~~~~,
$$
with $k_1 \ge k_2 \ge \dots \ge k_r > 0$, and
${\overline k}_1 \ge {\overline k}_2 \ge \dots
\ge {\overline k}_s > 0$,
coincide with the particle-hole excitations obtained from
$|\Delta N , d \rangle_0$.
Using the expressions of $\v_0^0$ and $\vb_0^0$ given in
\eq{fockw}, one finds that
\bea
\v_0^0 ~|\Delta N , d ; \{k_i\},\{{\overline k}_j\}
\rangle_0&=& \left(\frac{\Delta N}{2} \ + d  \right)
|\Delta N , d ; \{k_i\},\{{\overline k}_j\} \rangle_0~~~,
\nl
\vb_0^0 ~ |\Delta N , d ; \{k_i\},\{{\overline k}_j\}
\rangle_0 \ &=& \left(\frac{\Delta N}{2} \ - d  \right)
|\Delta N , d ; \{k_i\},\{{\overline k}_j\} \rangle_0~~~.
\nonumber
\ena
Thus, the bosonic field
built out of the $\v^0_\ell$ and $\vb_\ell^0$,
which describes the density fluctuations of the original
free fermions, is compactified on a circle of radius $r_0=1$.

Let us now consider the $1/N$-term of the effective
hamiltonian in \eq{hcv1}.
Due to the left-right mixing term proportional to $g$,
${\cal H}_{(1) }$ is not diagonal on the descendant states
$|\Delta N , d ; \{k_i\},\{{\overline k}_j\} \rangle_0$.
However, it can be diagonalized \cite{lutt} using
the Sugawara construction: replacing
$\left(\v_0^1 +\vb_0^1\right)$
with the expression given by \eq{mod1}, we obtain a quadratic
form in $\v_\ell^0$ and $\vb^0_{\ell}$,
which can be now diagonalized by means of the following
Bogoliubov transformation
\barr
\w^0_{\ell}&=&\v^0_{\ell}\ \cosh \beta + \vb^0_{-\ell}\
\sinh \beta ~~~, \nl
\wb^0_{\ell}&=&\v^0_{-\ell}\ \sinh \beta +
\vb^0_{\ell}\ \cosh \beta
\label{bogo}
\earr
for all $\ell$, with $ \tanh 2\beta =g/(2+g)$.

The operators
$\w^0_{\ell}$ and $\wb^0_{\ell}$
satisfy the same abelian Kac-Moody algebra with
central charge $c=1$ as the original
operators $\v^0_{\ell}$ and $\vb^0_{\ell}$ (cf. \eq{walg0}).
Then, by means of the generalized Sugawara construction,
we can define a new realization of the $\winf$algebra whose
generators $\w_\ell^i$ and $\wb_\ell^i$ are forms of degree
$(i+1)$ in $\w^0_{\ell}$ and $\wb^0_{\ell}$ respectively.
Consequently, the effective hamiltonian,
up to order $1/N$, becomes
\cite{flsz}
\bea
{\cal H}_{(1/N)}&\equiv& \left(2\pi \rho_0\right)^2
\left( {\cal H}_{(0)}+\frac{1}{N}\,{\cal H}_{(1)}\right)
\nl
&=& \left(
2\pi\rho_0  \sqrt{\lambda}\right)^2
\left[\left(\frac{\sqrt{\lambda}}{4}\,\w_0^0
+\frac{1}{N}\,\w_0^1\right)+
\left(\,W~\leftrightarrow~{\overline W}\,\right)
\right]~~~,
\label{h1n}
\ena
where
\beq
\lambda \equiv \exp(2\beta)= \sqrt{1+g}~~~.
\label{deflam}
\eeq
Notice that ${\cal H}_{(1/N)}$ is not diagonal on
$|\Delta N , d ; \{k_i\},\{{\overline k}_j\} \rangle_0$,
because the highest weight states of the
new algebra do not coincide with the
vectors $|\Delta N , d \rangle_0$, as is clear from \eq{bogo}.
However, the Bogoliubov transformation does not mix states
belonging to different Verma moduli.
This implies that the new highest weight vectors,
$|\Delta N ; d \rangle_W$, are still
characterized by the numbers $\Delta N$ and $d$ with the
same meaning as before, but their charges are different.
More precisely
\bea
\w_0^0 ~|\Delta N ; d \rangle_W  &=&
\left(\sqrt{\lambda}\,\frac{\Delta N}{2}+
\frac{d}{\sqrt{\lambda}}\right)
|\Delta N ; d \rangle_W \nl
\wb_0^0 ~|\Delta N ; d \rangle_W  &=&
\left(\sqrt{\lambda}\,\frac{\Delta N}{2}-
\frac{d}{\sqrt{\lambda}}\right)
|\Delta N ; d \rangle_W~~~.
\label{vdo}
\ena
{}From the last two equations we deduce that $\w^0_\ell$
and $\wb_\ell^0$ are still the modes of a compactified
bosonic field, but on a circle of radius
$r=1/\sqrt{\lambda}=\exp(-\beta)$.
This new field describes the density fluctuations
of the {\it interacting} fermions.

The highest weight states $|\Delta N , d \rangle_W$
together with their descendants, which we denote by
$|\Delta N , d ; \{k_i\},\{{\overline k}_j\} \rangle_W$,
form a {\it new} bosonic basis for our
theory that has no simple expression in
terms of the original free fermionic degrees of freedom.
The main property of this new basis is that it diagonalizes
the effective hamiltonian up to order $1/N$.
In fact, using Eqs. (\ref{h1n}) and (\ref{vdo}),
it is easy to check that
$$
{\cal H}_{(1/N)}
{}~|\Delta N , d ; \{k_i\},\{{\overline k}_j\} \rangle_W =
{\cal E}_{(1/N)}
{}~|\Delta N , d ; \{k_i\},\{{\overline k}_j\} \rangle_W
$$
where \cite{kaya}
\beq
{\cal E}_{(1/N)} =  \left(
2\pi\rho_0  \sqrt{\lambda}\right)^2\Bigg[\frac{\lambda}{4}\,\Delta N
+
\frac{1}{N}\left(\lambda\,\frac{\left(\Delta N\right)^2}{4}  +
\frac{d^2}{\lambda} +
k+{\overline k} \right)\Bigg]
{}~~~,
\label{fsize}
\eeq
with $k=\sum_i k_i$ and
${\overline k}=\sum_j {\overline k}_j~$.
These eigenvalues are clearly degenerate when $k \geq 2$ or
${\overline k} \geq 2$.

Since the effective
hamiltonian ${\cal H}$ has been derived within a perturbative
approach,
an expansion in the coupling constant $g$
should be understood in all previous formulae.
However, if we limit our analysis to the $1/N$-terms,
nothing prevents us from improving our results by
extending them to all orders in $g$. Indeed, the Bogoliubov
transformation (\ref{bogo}) diagonalizes the hamiltonian
${\cal H}_{(1)}$ {\it exactly}, and the resulting expression
depends on the coupling constant only through $\lambda$,
which contains all powers of $g$! This improvement is
a well-known result in the Luttinger model \cite{lutt,fermi},
but we would like to stress that in our case
it can be done only if we disregard
the $O(1/N^2)$-terms of the hamiltonian, because
the Bogoliubov transformation
(\ref{bogo}) does not diagonalize ${\cal H}_{(2)}$.

To investigate this issue, let us analyze the $1/N^2$-term
of the effective hamiltonian \eq{hcv2}. Using the generalized
Sugawara construction, Eqs. (\ref{mod0}--\ref{mod2}), we
first rewrite ${\cal H}_{(2)}$ as a cubic form in $\v^0_{\ell}$
and $\vb^0_{\ell}$, and then perform the Bogoliubov
transformation (\ref{bogo}) to re-express it in terms of
the $\w^i_\ell$ and $\wb^i_\ell$ generators.
A straightforward calculation leads to
${\cal H}_{(2)} = {\cal H}_{(2)}' \ +\ {\cal H}_{(2)}''$,
where
\beq
{\cal H}_{(2)}'=
\sqrt{\lambda} \left( \w^2_0+\wb^2_0 \right)-
\frac{\sqrt{\lambda^3}}{12}\left(\w_0^0+\wb_0^0\right)
- \frac{g}{2 \lambda}
\,\sum_{\ell=1}^{\infty}\,\ell\left(
\w^0_{-\ell}\,\w^0_{\ell}+ \wb^0_{-\ell}\,\wb^0_{\ell}
\right) ~~~,
\label{h2w'}
\eeq
and
\beq
{\cal H}_{(2)}'' \,=\, -\,\frac{g}{2 \lambda}
\,\sum_{\ell=1}^{\infty}\,\ell\left(
\w^0_{\ell}\,\wb^0_{\ell}+ \w^0_{-\ell}\,\wb^0_{-\ell}
\right)~~~.
\label{h2w''}
\eeq
Neither ${\cal H}_{(2)}'$ nor ${\cal H}_{(2)}''$ are diagonal
on the states
$|\Delta N , d ; \{k_i\},\{{\overline k}_j\} \rangle_W$
considered so far. In fact, these states are
not in general eigenstates of $\left(\w_0^2+\wb_0^2\right)$,
and hence cannot be eigenstates of
${\cal H}_{(2)}'$; moreover, since they have definite
values of $k$ and ${\overline k}$, they cannot be eigenstates
of ${\cal H}_{(2)}''$ either, because this operator
mixes the left and right sectors.

It is not difficult, however, to overcome
these problems. Since
${\cal H}_{(2)}'$ and ${\cal H}_{(1/N)}$ commute with each
other, it is always possible to find suitable
combinations of the states
$|\Delta N , d ; \{k_i\},\{{\overline k}_j\} \rangle_W$
with fixed $k$ and ${\overline k}$ that diagonalize
simultaneously ${\cal H}_{(2)}'$ and ${\cal H}_{(1/N)}$,
therefore lifting the degeneracy of the
spectrum present to order $1/N$.
The term ${\cal H}_{(2)}''$, instead, has to be treated
perturbatively, but only to first order in $g$.
In fact to higher orders, the spurious states
introduced extending to infinity the sum in \eq{kindue}
would contribute as intermediate states.
These contributions, however, would be meaningless because
the hamiltonian to order $O(1/N^2)$ is not even bounded below.
{}From \eq{h2w''} it is easy to check that ${\cal H}_{(2)}''$
has vanishing expectation value on any state that is
simultaneously eigenstate of ${\cal H}_{(1/N)}$ and
${\cal H}_{(2)}'$.
Thus, according to (non-degenerate) perturbation theory,
${\cal H}_{(2)}''$ has no effect on the energy spectrum
to first order in $g$.

In view of these considerations, we neglect
${\cal H}_{(2)}''$ and regard as the effective
hamiltonian the following operator
\bea
{\cal H} &\equiv&
{\cal H}_{(1/N)} + \left(
2\pi\rho_0 \right)^2
\frac{1}{N^2}\,{\cal H}_{(2)}'\nl
&=&\left(
2\pi\rho_0  \sqrt{\lambda}\right)^2
\left\{\left[\frac{\sqrt{\lambda}}{4}\,\w_0^0
+\frac{1}{N}\,\w_0^1+
\frac{1}{N^2}\left(\frac{1}{\sqrt{\lambda}}\,\w_0^2
-\frac{\sqrt{\lambda}}{12}\,\w_0^0 \right.\right.
\right.\nl
&& -\left.\left.\left.
\frac{g}{2\lambda^2}\,\sum_{\ell=1}^\infty
\,\ell~\w_{-\ell}^0\,\w_\ell^0\right)
\right]+\left(\,W~\leftrightarrow~{\overline W}\,\right)
\right\}~~~.
\label{hcsf}
\ena
Obviously, to be consistent with our perturbative approach,
we should keep in the r.h.s. of \eq{hcsf} only the terms
that are linear in $g$.

We now compare the eigenvalues of
${\cal H}$ to the exact low-energy spectrum of the
Calogero-Sutherland model obtained from the Bethe Ansatz
method \cite{kaya}.
Any low-energy solution of the Bethe Ansatz equations
is labeled by a set of integer numbers
\beq
I_j\ =\ \frac{2j-1-N'}{2} + d - {\overline n}_j +  n_{N'-j+1}
\label{Ii}
\eeq
where $N'= N + \Delta N$, and $j=1, \dots, N'$.
The integers $n_j$ are ordered according to
$n_1\geq n_2\geq \dots \geq 0$
and are different from zero only if $j < \Lambda \ll N$ (and
analogously
for ${\overline n}_j$).

By generalizing to order $1/N^2$ the procedure presented in
Ref. \cite{kaya}, we have derived \cite{flsz} the exact energy of
the excitation described by the numbers (\ref{Ii}):
\bea
\tilde{\cal E} &=& \left(2\pi\rho_0 \sqrt{\xi}\right)^2
\Bigg\{\Bigg[\frac{\sqrt{\xi}}{4}\,Q+\frac{1}{N}
\Bigg(\frac{1}{2}\,Q^2+ \sum_j n_j\Bigg)
\nl
&&+ \frac{1}{N^2}\Bigg(\frac{1}{3\sqrt{\xi}}\,Q^3
-\frac{\sqrt{\xi}}{12}\,Q +
\frac{2 \sum_j n_j}{\sqrt{\xi}}\,Q + \frac{\sum_j n_j^2}{\xi}
-\sum_j \left(2j-1\right) n_j\Bigg)\Bigg]
\nl
&& +\left(Q\, \leftrightarrow \, {\overline Q}~,
{}~\{n_j\} \, \leftrightarrow \,
\{{\overline n}_j\} \right)\Bigg\}~~~,
\label{eba}
\ena
where
\beq
\xi=\frac{1+\sqrt{1+2g}}{2}~~~,
\label{xi}
\eeq
and
\beq
Q=\sqrt{\xi}\,\frac{\Delta N}{2}+
\frac{d}{\sqrt{\xi}}~~~~,~~~~
{\overline Q}=\sqrt{\xi}\,\frac{\Delta N}{2}-
\frac{d}{\sqrt{\xi}}~~~.
\label{Q}
\eeq

Of course, being an exact result, \eq{eba} holds to all
orders in $g$. Comparing Eqs. (\ref{deflam}) and (\ref{xi}),
we see that
\beq
\xi=\lambda+O(g^2)~~~.
\label{xilam}
\eeq
Comparing \eq{eba} with \eq{fsize}, we realize that, at least to
order $1/N$, the exact results can be obtained from the perturbative
ones simply by changing $\lambda$ into $\xi$.
Moreover, in Ref. \cite{flsz} we have checked on several explicit
examples
that the eigenvalues of ${\cal H}$ with $\xi$ in place
of $\lambda$ coincide with the exact energy of the
low-lying excitations given by \eq{eba}.
Thus, we are led to
conjecture that the {\it exact} effective
hamiltonian of the
Calogero-Sutherland model is given by
\eq{hcsf} with $\xi$ in place of $\lambda$.
We may consider this operator as a non-perturbative improvement
of ${\cal H}$ which was derived in perturbation theory.

We conclude by mentioning that our method of
algebraic bosonization can be applied in principle to any
gapless fermionic hamiltonian consisting of a
bilinear kinetic term and an arbitrary four-fermion interaction.
No special requirements on the form of the dispersion relation
and the potential are needed.
In particular, it is not necessary for the system to be integrable.
In Ref. \cite{flsz} we have also discussed the algebraic
bosonization of the Heisenberg model, by mapping it into a
theory of fermions on a lattice by means of a Jordan-Wigner
transformation.

By comparison with the Bethe Ansatz solution it can be shown
\cite{ms} that also the one-dimensional Bose gas, with hamiltonian
\beq
H = \int_0^L dx \left[ \partial_x \phi^{\dagger}(x)\, \partial_x
\phi(x)
+ 2 \kappa \ \phi^{\dagger}(x)\,\phi^{\dagger}(x)\,\phi(x)\,\phi(x)
\right]~~~,
\label{hb}
\eeq
with $\kappa>0$, has a spectrum of low-energy excitations that
can be described by the representation theory of the
$\winf\times {\overline \winf}$algebra.
In particular, to first order in a $1/{\kappa}$-expansion, the effective
hamiltonian corresponding to \eq{hb} turns our to be
\bea
{\cal H} &=& \left(2\pi\rho_0 \right)^2
\left\{\frac{1}{4} \left(1-\frac{5}{3}g \right)\,\w_0^0
+\frac{1}{N} \left(1- 2g \right)\,\w_0^1 \right.
\nl
&& +\frac{1}{N^2}\left[\left(1- 2g \right)\,\w_0^2
-\frac{1-3g}{12}\,\w_0^0 - 2g \w_0^0\,\wb_0^0\right]
\\ \label{hhb}
&& \left. - \frac{2g}{N^3} \left[\left(\w_0^2 + \wb_0^2\right)
\w_0^0 -\frac{1}{2} \left(\w_0^1 - \wb_0^1\right)^2
-\frac{1}{24} \left(\w_0^0 + \wb_0^0\right)^2 \right]
+\left(\,W~\leftrightarrow~{\overline W}\,\right) \right\}
\nonumber
\ena
where $\rho_0=N/L$ and $g=2\rho_0/{\kappa}$.
Turning to the fermionic realization of the
$\winf\times {\overline \winf}$algebra, one can calculate
the first order correction to the gas of free fermions, which
is equivalent to the gas of one-dimensional boson
($\kappa=\infty$),
and find the following hamiltonian
$$
H = - \int_0^L dx ~ \psi^{\dagger}(x) \,\partial_x^2 \psi(x) +
\frac{2}{\kappa L} \int_0^L dx\!\int_0^L dy~
\psi^{\dagger}(x)\,\psi^{\dagger}(y) \Big(\partial_x - \partial_y
\Big)^2
\psi(x)\,\psi(y)~~~.
$$

\end{document}